\documentclass[reprint,amsmath,amssymb,aps,twocolumn,prb,
floatfix,
noeprint
]{revtex4-2}

\usepackage{amsmath}
\usepackage{amssymb}
\usepackage{graphicx}
\usepackage{dcolumn}
\usepackage{bm,array}
\usepackage[T1]{fontenc}
\usepackage{url}
\begin{document}
\preprint{APS/123-QED}

\title{Magnetic Field Controlled Surface Localization\\of Spin-Wave Ferromagnetic Resonance Modes in 3D Nanostructures}

\author{Mateusz Gołębiewski}
\email{mateusz.golebiewski@amu.edu.pl}
\affiliation{Institute of Spintronics and Quantum Information, Faculty of Physics, Adam Mickiewicz University, Uniwersytetu Poznańskiego 2, 61-614 Poznań, Poland}

\author{Krzysztof Szulc}
\affiliation{Institute of Spintronics and Quantum Information, Faculty of Physics, Adam Mickiewicz University, Uniwersytetu Poznańskiego 2, 61-614 Poznań, Poland}

\author{Maciej Krawczyk}
\affiliation{Institute of Spintronics and Quantum Information, Faculty of Physics, Adam Mickiewicz University, Uniwersytetu Poznańskiego 2, 61-614 Poznań, Poland}

\date{\today}

\begin{abstract}
Extending the current understanding and use of magnonics beyond conventional planar systems, we demonstrate surface localization of spin-wave ferromagnetic resonance (FMR) modes by designing complex three-dimensional nanostructures. Using micromagnetic simulations, we systematically investigate woodpile-like scaffolds and gyroids -- periodic chiral entities characterized by their triple junctions. The study highlights the critical role of demagnetizing fields and exchange energy in determining the FMR responses of 3D nanosystems, especially the strongly asymmetric distribution of the spin-wave mode over the system height. Importantly, the top--bottom dynamic switching of the surface mode localization across the structures in response to changes in magnetic field orientation provides a new method for controlling magnetization dynamics. The results demonstrate the critical role of the geometric features in dictating the dynamic magnetic behavior of three-dimensional nanostructures, paving the way for both experimental exploration and practical advances in 3D magnonics.
\end{abstract}

\maketitle

\section{Introduction \label{Sec:int}}
Spin waves (SWs), originating from the collective oscillations of magnetic moments, are marked by their intricate dynamics, dependence on material structure, magnetization texture, and profound application potential in IT systems~\cite{Dieny2020}. The properties of SWs are a consequence of the interplay between long-range magnetostatic and short-range exchange interactions. This balance is particularly pronounced when the anisotropy of the magnetostatic interactions introduces a unique dependence of the SW propagation on the alignment between magnetization and SW wavevector. Such dependencies give rise to a number of distinctive SW properties, including negative group velocity, caustics, pronounced nonlinearity, and dynamic reconfigurability~\cite{Pirro2021AdvancesMagnonics}. Thus, the current focus of the magnonic research is not only to understand these phenomena, but also to exploit them for digital, analog, non-conventional, and quantum signal processing at high frequencies, from a few to hundreds of GHz, operating at the nanoscale, and consuming less power than other alternative systems~\cite{Chumak2019}. This vision is reflected in recent breakthroughs and strategic roadmaps of the field~\cite{Barman2021TheRoadmap, Chumak2022AdvancesComputing}.

Nanostructured 3D networks can significantly advance the potential of magnonics, giving rise to topological and geometric effects and emergent material properties that extend existing and offer new possibilities for SW manipulation~\cite{Gubbiotti2019Three-DimensionalMagnonics,hertel_defect-sensitive_2022, Krawczyk2014Review}. For example, by tuning their geometry and lattice period, we obtain control over the magnonic band structure with tailored SW group velocity and collective response to external stimuli in all possible directions of SW propagation and polarization of external fields~\cite{Krawczyk2008,Krawczyk2014Review,Okuda_2017}. In addition, 3D magnetic structures provide an opportunity to explore chiral surface, edge, and corner states, characterized by higher-order topology~\cite{Kondo2019,Hua2023}. Furthermore, a fully interconnected 3D systems~\cite{Fischer2020LaunchingNanostructures, Makarov2022NewNanoarchitectures, Cheenikundil2022High-frequencyNanoarchitecture} open up a new degree of freedom to explore other emerging phenomena in magnetism, such as frustration and magnetic charge isolation, realized with 3D artificial spin-ice systems~\cite{May2021MagneticSpin-ice, Saccone2023}. In recent years, the significant development of new fabrication techniques---such as two-photon lithography, focused-electron-beam deposition, and block-copolymer templating---makes it possible to fabricate and measure complex 3D structures and artificial nanosystems on the nanometer scale~\cite{Llandro2020VisualizingNetworks, Fernandez-Pacheco2017Three-dimensionalNanomagnetism, Donnelly2022ComplexNanostructures, Hunt2020HarnessingNanoscale, vandenBerg2023CombiningNanostructures, Yan2012EvaluationsMelting}, test them for various applications~\cite{Yanez2016CompressiveApplications, Turner2013MiniatureCrystals}, and also start investigations on SW dynamics~\cite{Grundler2023}. However, the study of magnetization dynamics in nanostructures with periodicity in 3D is still in its early stages~\cite{May2021MagneticSpin-ice, Grundler2023}.

\begin{figure*}[t!]
\includegraphics[width=\linewidth]{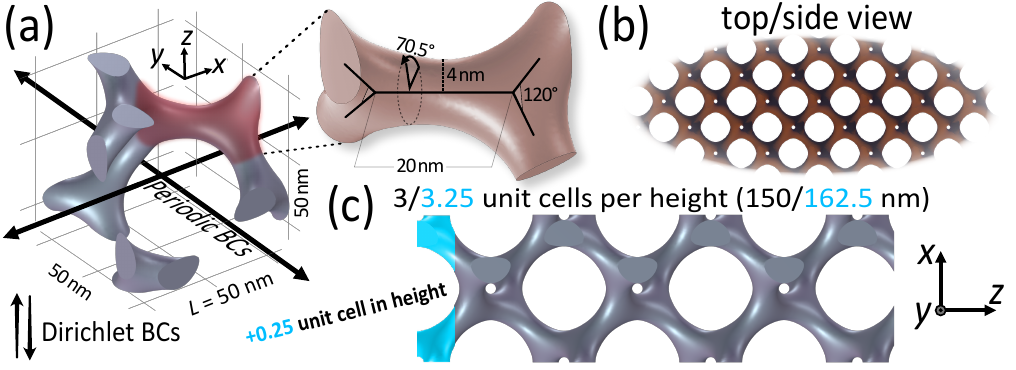}
\caption{Visualization of gyroidal systems used in micromagnetic simulations. The cubic gyroidal unit cell (UC) with dimensions $L=50$~nm and a volume fraction of $\phi=10\%$ is shown in (a), with the enlarged inset highlighting the chiral linkage between two primary gyroidal nodes. The arrows indicate the direction of application of the periodic boundary conditions (PBCs), which are applied along both the $x$- and $y$-axes defining the plane of the structure. Along the $z$-axis -- perpendicular and far from the plane -- Dirichlet BCs are assumed [see Methods~(\ref{Sec:sim})]. Panel (b) shows the orthographic projection of the gyroid structure from each side of the analyzed cubic cells (crystallographic normal direction [100]), demonstrating the characteristic square distribution of gyroid channels, rotated 45~deg to the axes. The analyses were performed using models with 3 and 3.25~UCs per height (c), highlighting the influence of the surface shape on the observed effects.
\label{Fig:gyr_str}}
\end{figure*}

Gyroids are 3D structures with intriguing properties that have been widely explored in photonics~\cite{saba2011,Flavell2023,Peng2016}. As detailed in Refs.~\cite{Schoen1970InfiniteSelf-intersections,Schoen2012reflections, Rosi2020hal}, the gyroid emerges as a unique triply-periodic minimal surface. Its defining feature is a zero mean curvature, which means that every point on the surface acts as a saddle point, characterized by equal and opposite principal curvatures~\cite{Dacorogna2014calculus}. Its intricate design comprises cubic unit cells (UCs) linked by nanorods with elliptical cross-section~\cite{Dolan2015OpticalMetamaterials}. The inherent chirality and curvature of nanoscale gyroids~\cite{Wohlgemuth2001triply} offer a promising avenue also for controlling non-collinear spin textures in magnetism~\cite{Lich2023FormationNanostructures}. This potential is further underscored by the visualization of magnetic structures in $\text{Ni}_{75}\text{Fe}_{25}$ gyroid networks~\cite{Llandro2020VisualizingNetworks}. Its chiral structure, well below 100~nm UCs, and the building struts' dimensions close to the exchange length also promise interesting SW dynamics. The effect of Ni gyroid crystallography on resonance frequencies has recently been demonstrated using micromagnetic simulations and broadband ferromagnetic resonance measurements~\cite{golebiewski2024gyr}.

In this paper, we study the magnetization dynamics in a thin film made of a 3D ferromagnetic gyroid nanostructure by theoretical analysis of ferromagnetic resonance (FMR) modes. Unexpectedly, we find that for some specific orientations of the in-plane bias magnetic field, the lowest frequency signal with the highest FMR intensity comes from a surface-localized mode. Interestingly, the localized mode is not propagating (wavevector $k=0$~rad/m), indicating that it is not a Damon--Eshbach type of localization~\cite{Damon1961}. It is also not a Shockley~\cite{Rychly2017} or topologically protected type of surface state~\cite{McClarty2022TopologicalReview}, which requires a Bragg bandgap in the SW spectrum. To explain this unusual type of localization, we use a simpler structure with vertically layered, orthogonally alternating cylindrical ferromagnetic nanorods. On this basis, we also reject the hypothesis that this localization is just due to the non-uniformity of the demagnetizing field at the surface of the structure (i.e., edge modes~\cite{Guo2013}), but we show that the demagnetizing field plays an important role by creating potential barriers for SWs, whose localization is determined by the exchange interaction. Thus, these results provide a new type of SW surface localization with a high absorption intensity of a homogeneous microwave field. Importantly, the localization can be controlled by rotation of the in-plane bias magnetic field as well as by shape manipulation, especially of the surface region of the structures. It allows the SW intensity to be transferred from the bottom surface, through the bulk, to the top surface by simply rotating the sample or the magnetic-field direction.

\section{Gyroid structure and numerical simulations}
The gyroidal surface divides space into two contrasting labyrinths that intersect at angles of 70.5~deg [Fig.~\ref{Fig:gyr_str}(a)], creating a captivating geometric pattern. Its representation can be expressed by the trigonometric equation:
\begin{equation}
\begin{split}
    \sin{(2\pi x/L)}\cos{(2\pi y/L)}&+\\ 
    \sin{(2\pi y/L)}\cos{(2\pi z/L)}&+\\ 
    \sin{(2\pi z/L)}\cos{(2\pi x/L)}&\le(101.5-2\phi)/68.1,
\end{split}
\label{eq:gyr_tryg}
\end{equation}
where $L$ signifies the gyroid UC length and $\phi$ is a filling factor.
In our research, the UC of the nickel-made gyroid measures 50~nm, as shown in Fig.~\ref{Fig:gyr_str}(a), which results in a single strut diameter of 8~nm, thus comparable to the exchange length~\cite{Llandro2020VisualizingNetworks}. It relates to the volume fraction of $\phi=10\%$. In this work, we study gyroids in the form of a thin films with the [100]-direction normal to the plane [Fig.~\ref{Fig:gyr_str}(b)] and with two different heights: 3~UCs (150~nm) and 3.25~UCs (162.5~nm) [Fig.~\ref{Fig:gyr_str}(c)].

To study the SW dynamics, we numerically solve an eigenproblem obtained from the Landau-Lifshitz (LL) equation in linear approximation. Using PBCs on the UC boundaries along the $x$- and $y$-axes, we model infinite gyroidal films (see Fig.~\ref{Fig:gyr_str}). Details on the technical and theoretical aspects of the simulations can be found in the Methods, Sec.~\ref{Sec:sim}. For each of our simulations, we adopt material parameters typical of nickel (Ni) films: the saturation magnetization $M_\text{s}=480$~kA/m, the exchange stiffness $A_\text{ex}=13$~pJ/m, and the gyromagnetic ratio $\gamma=176$~rad/s/T~\cite{Coey2010MagnetismMaterials,Singh1976CalculationNickel}. The external magnetic field, $\mu_0 H_\text{ext}$, remains constant at 500~mT. This magnitude, as supported by Ref.~\cite{Llandro2020VisualizingNetworks}, validates our assumption of complete saturation of magnetization of the gyroid structure.
 
\section{Results}

\subsection{Surface modes in gyroid\label{sec:h_rot}}
\begin{figure}[htp]
\includegraphics[width=\linewidth]{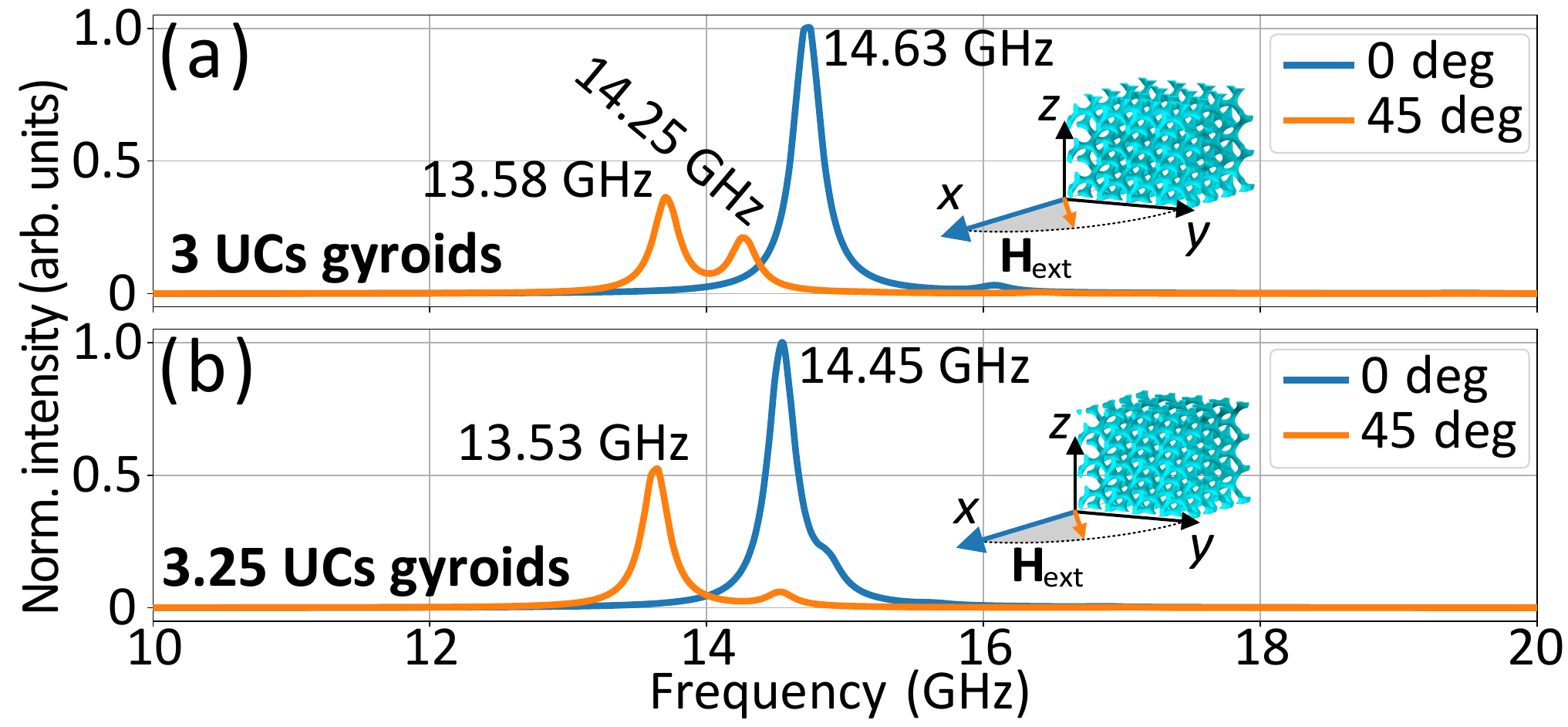}
\caption{Normalized resonance spectra of the gyroid structures with 3~UCs (a) and 3.25~UCs (b) per height. The colors indicate various angles of the external magnetic field relative to the $x$-axis. The values of the corresponding frequencies are given for the high-intensity peaks.
\label{Fig:gyr_fmr}}
\end{figure}

\begin{figure}[htp]
\includegraphics[width=\linewidth]{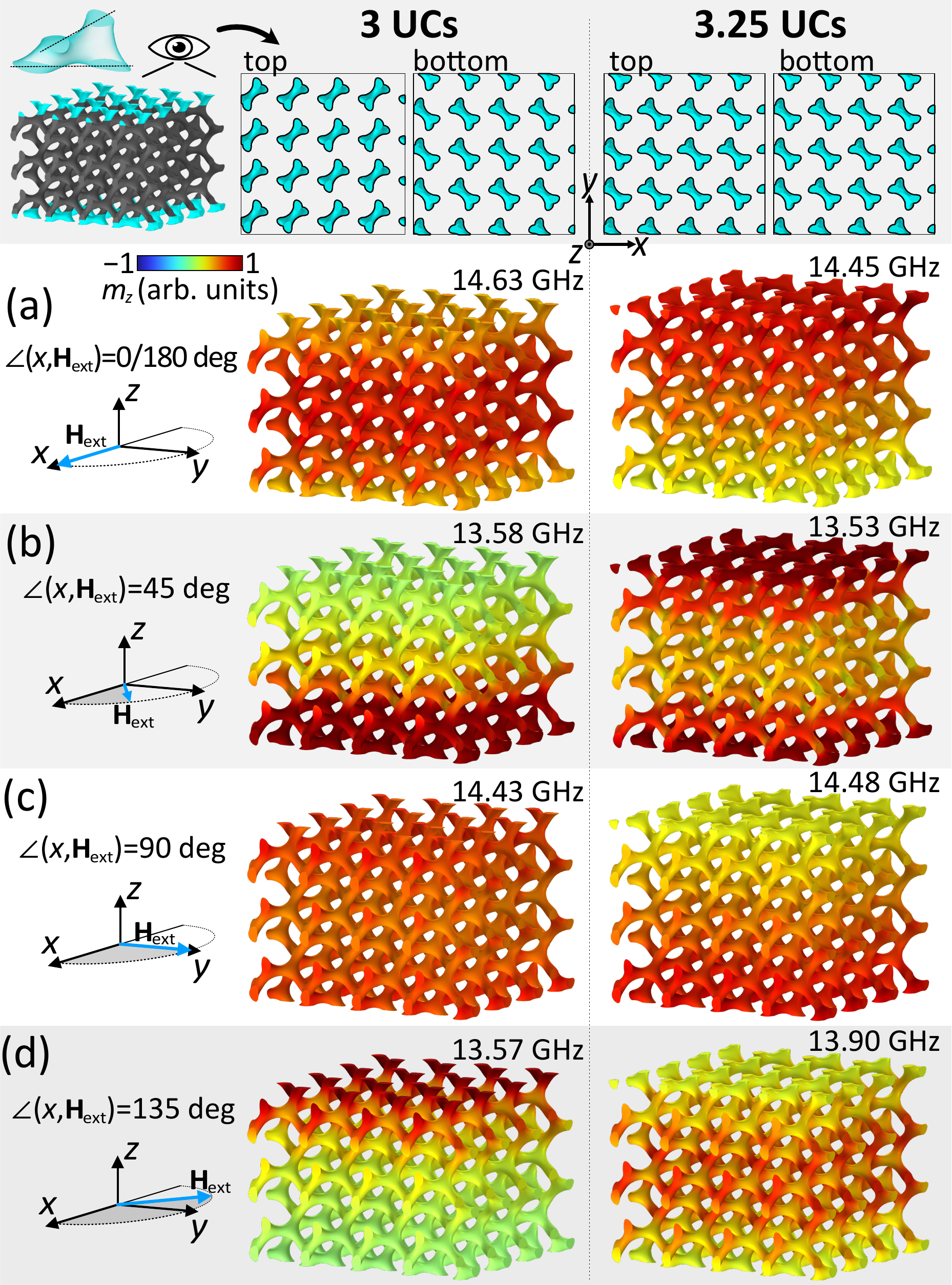}
\caption{Distribution of the dynamic component of the magnetization $m_z$ in a gyroid structure with a $\phi=10\%$ and height of 150~nm (3~UCs -- angled surface struts on the top and bottom surface, left side) and 162.5~nm (3.25~UCs, parallel surface struts, right side). Above are the orientations of the outer struts for the two configurations studied (top view). Different configurations of the direction of the external magnetic field in the plane of the layer are shown, demonstrating the differences in the localization of individual FMR modes. For better visibility, the gyroidal arrays consist of $4\times4$ columns of the UC.
\label{Fig:gyr_loc}}
\end{figure}

We examine the FMR response of the gyroid structure as a function of the external magnetic field (500~mT) direction within its $xy$-plane. As mentioned above, we consider films with two heights, 150~nm (3 UCs) and 162.5~nm (3.25 UCs), which form angled and parallel patterns on the top and bottom surfaces of the gyroid film, respectively (see the top panels in Fig.~\ref{Fig:gyr_loc} showing the struts at the top and bottom of the film as viewed from above). As shown in Fig.~\ref{Fig:gyr_fmr}, the FMR spectra [for 3 UCs (a) and 3.25 UCs (b)] for 0~deg angle of static field orientation consist of only a single peak of high intensity [all spectra intensities were calculated using Eq.~(\ref{eq:int}) based on the response to the homogeneous microwave magnetic field excitation $h_\text{dyn}$ -- see Methods, Sec.~\ref{Sec:sim}]. In addition to the frequency decrease, when the field is rotated by 45~deg, the spectra in Fig.~\ref{Fig:gyr_fmr} also clearly show the emergence of a secondary peak of lower intensity. The lowest-frequency, high-intensity peaks are attributed to the modes with in-phase magnetization oscillations over the height, while the secondary peaks are attributed to an asymmetric quantized SW mode [see Fig.~S1(a) and (b) in the Supplementary Information]. In the following, we will focus only on the resonant frequency mode, which has the most intense response to the homogeneous microwave magnetic field and is the lowest frequency mode in a given configuration.

Unexpectedly, the rotation of the magnetic field along with the orientation of the outer parts of the gyroid affects not only the frequency but also the amplitude distribution of this SW mode, as shown in Fig.~\ref{Fig:gyr_loc}. Specifically, when the field aligns with the $x$-axis for 3~UCs case [Fig.~\ref{Fig:gyr_loc}(a)], magnetization distribution appears nearly uniform across the layer's entire height, exhibiting a bulk concentration preference, henceforth referred to as a bulk mode (frequency 14.63~GHz). However, for a 3.25~UCs, i.e., where the top and bottom of the structure are oriented the same, we can see the distribution of the magnetization tendency towards localization in the upper level (14.45~GHz). A significant change in amplitude distribution occurs when the field is rotated by 45~deg -- the FMR mode is predominantly concentrated in one or both of the surface regions of gyroids for the 3~UCs and 3.25~UCs cases (13.58 and 13.53~GHz), respectively, which we refer to as surface modes. A subsequent rotation by another 45~deg aligns the field along the $y$-axis [Fig.~\ref{Fig:gyr_loc}(c)], transitioning the mode back to an almost uniform state for 3~UCs structure (14.43~GHz), albeit with a distinct bias towards both surfaces. For the 3.25~UCs case (14.48~GHz), we again see the tendency of the magnetization to localize, but this time on the bottom of the layer. At 135~deg rotation [Fig.~\ref{Fig:gyr_loc}(d)], the pattern of strong localization reemerges for 3~UCs structure (13.57~GHz), yet on the opposite surface compared with 45~deg, showcasing a dynamic shift in vertical localization of the FMR mode depending on the field's rotation. At the same time, for a 3.25~UCs structure (13.90~GHz), the field directed at an angle of 135~deg from the $x$-axis in the plane, causes the magnetization to be concentrated in the inner, bulk part of the structure. Completing the cycle, a 180~deg rotation reinstates the magnetization distribution to its original states observed at 0~deg in both cases.

Gyroidal systems, due to their complexity, significantly complicate the interpretation of the results obtained, therefore to gain a deeper insight into the mode localization phenomena, we propose a model of a simple three-dimensional structure with reduced complexity. As a result of the systematic study (see Supplementary Information, Fig.~S6), a woodpile-like scaffold structure, in which the horizontal piles are separated and connected with vertical bars, emerged as an optimal candidate that meets the criteria, notably:
\begin{itemize}
    \item material continuity essential for facilitating exchange interactions,
    \item alternating and perpendicular configuration of nanorods designed to influence the modulation of demagnetizing field distribution,
    \item a size and spacing between nanorods aligning with the order of magnitude of the exchange length $l_\mathrm{ex}=\sqrt{2A_\mathrm{ex}/(\mu_0M_\mathrm{s}^2)}$($\approx9.5$~nm for the Ni parameters).
\end{itemize}
As we will show, it captures the essential geometric attributes necessary to replicate the magnetic field angle-dependent localization effects observed in gyroids.

\subsection{Woodpile-like scaffolds \label{sec:ladd}}

\begin{figure}[htp]
\includegraphics[width=\linewidth]{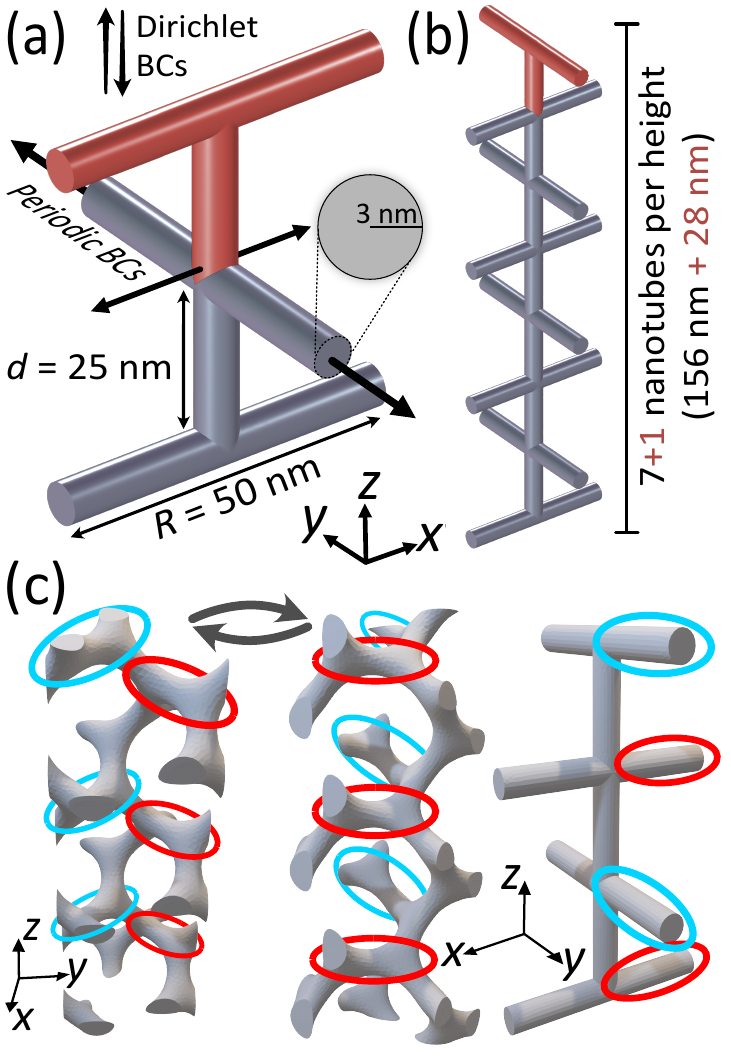}
\caption{Visualization of the scaffold models used in the micromagnetic simulations. The enlarged representative cutout in (a) illustrates the main geometric features of the nanorods, including their length $R=50$~nm, vertical distance $d=25$~nm, and circular cross-section radius $r=3$~nm. The arrows indicate the direction of the applied boundary conditions, in the same way as for gyroids (see Fig.~\ref{Fig:gyr_str}). The red color marks an additional level used to manipulate the vertical symmetry of the whole structure. Model (b) shows a full-height column used to perform calculations with a symmetric arrangement (with 7 horizontal nanorods) and an asymmetric arrangement with 8 nanorods, where the bottom one is rotated 90~deg with respect to the top one. In (c), the critical gyroid rods for analysis are highlighted, schematically illustrating the structural similarities between the two systems studied. Blue lines indicate the hard axis struts, while red circles mark the easy axis struts, which act as energy "barriers" under the influence of an external magnetic field oriented parallel to them.
\label{Fig:ladd_str}}
\end{figure}

The proposed woodpile-like scaffold structure is a stack of vertically and orthogonally distributed cylindrical nanorods, the UC of which is shown in Fig.~\ref{Fig:ladd_str}(a), with the vertical distance between them defined as $d$. In the micromagnetic simulations, the radius of cylinders was kept constant at $r=3$~nm and the width of the cubic UC (nanorod length) at $R=50$~nm, with PBC in the $xy$-plane, as in the case of gyroids. The selection of this structural type is driven by the analogies observed in the distribution of critical struts. As depicted in Fig.~\ref{Fig:ladd_str}(c), the hard and easy axis struts are arranged quasi-perpendicularly, functioning as energy barriers under the influence of an external magnetic field in both systems. In addition, the scaffold nanostructures allow fast and efficient analysis of the FMR mode distribution, taking into account an even number of orthogonal rods (asymmetry through height -- surface top/bottom bars are perpendicular to each other as in the 3-UCs gyroid structure) and an odd number of them (symmetry -- surface top/bottom bars are parallel to each other, as in the 3.25-UCs gyroid structure), as shown in Fig.~\ref{Fig:ladd_str}(b). Thus, according to the framework established in Sec.~\ref{sec:h_rot}, the scaffold structure has been categorized into two configurations: symmetric, with an odd number of scaffold levels, and asymmetric, with an even number of levels. In our work, the symmetric configuration is represented by the 156~nm-high structure consisting of 7 levels of nanorods (7~UCs), while the asymmetric configuration measures 184~nm in height with 8 levels of nanorods (8~UCs). Unlike gyroids, the entire structure, not just the surface reconstructions, has inversion symmetry.

\begin{figure}[htp]
\includegraphics[width=\linewidth]{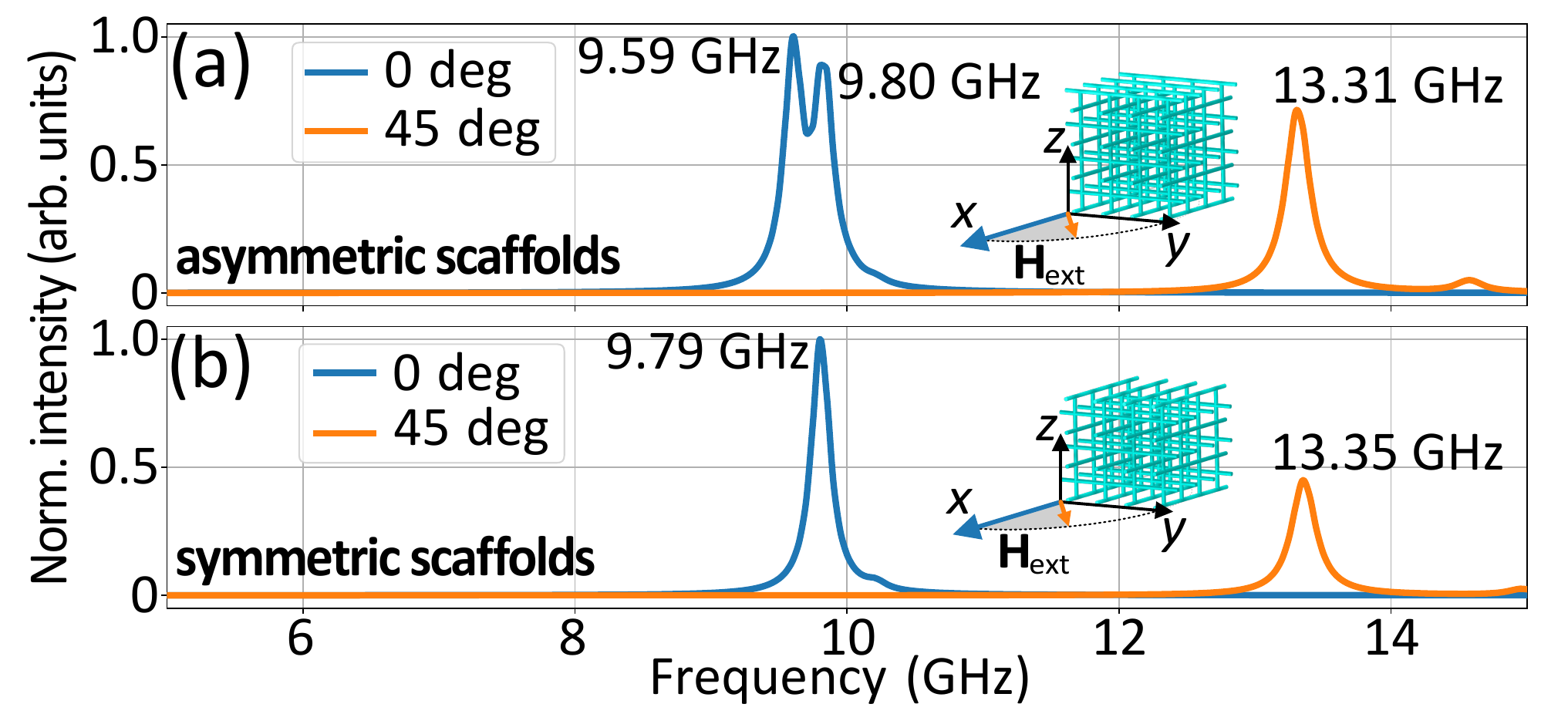}
\caption{Normalized resonance spectra of the woodpile-like scaffold structures used in the study -- asymmetric with 8 horizontal nanorods (a) and symmetric with 7 horizontal nanorods (b) per height. The colors indicate various angles of the external magnetic field relative to the $x$-axis. The values of the corresponding frequencies are given for each high-intensity peak.
\label{Fig:scaff_fmr}}
\end{figure}

\begin{figure}[htp]
\includegraphics[width=\linewidth]{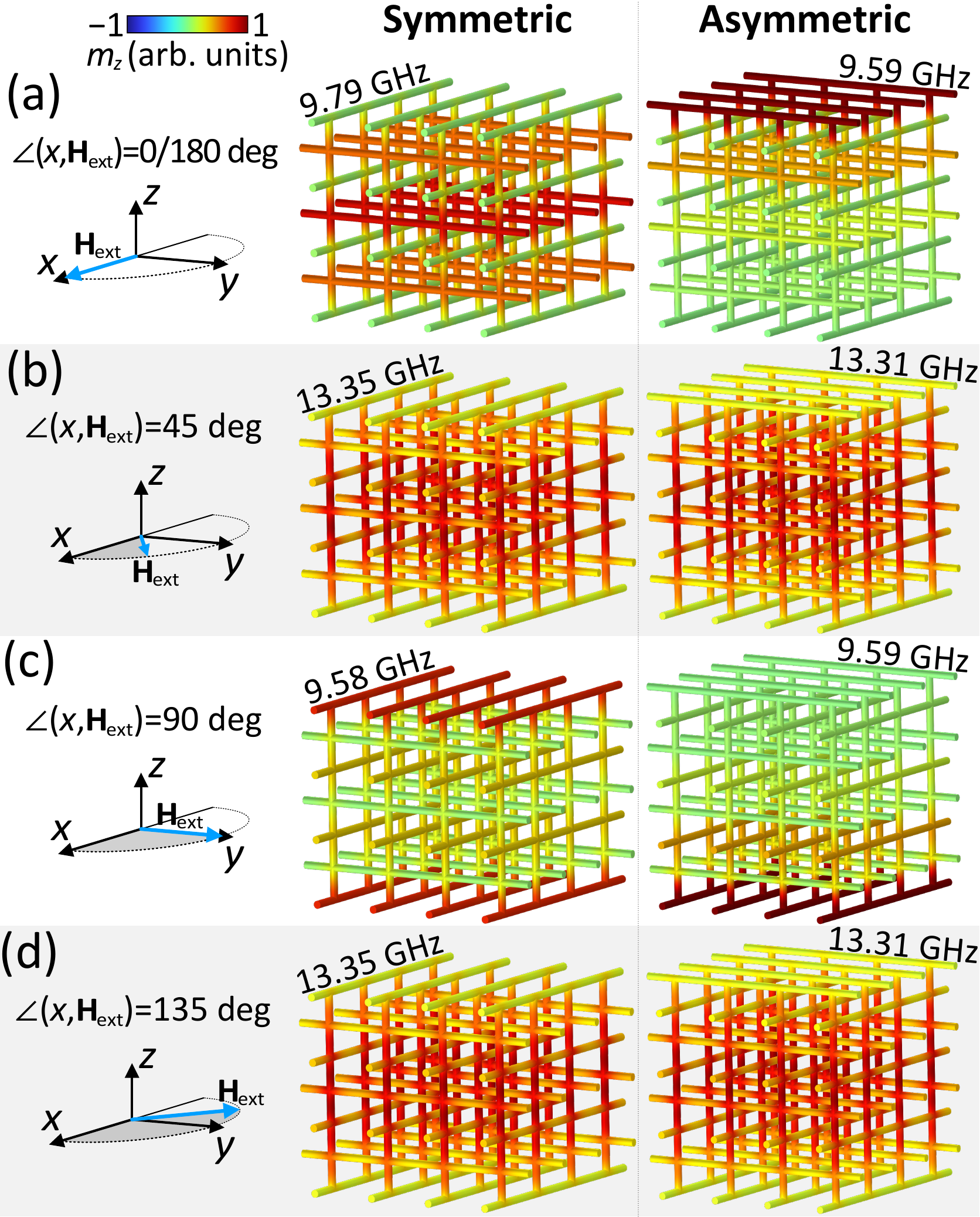}
\caption{The dynamic magnetization component distribution, $m_z$, across scaffold structures featuring symmetric (156~nm height) and asymmetric (184~nm height) configurations, subjected to different orientations of the external magnetic field within the plane, adjusted in 45-deg increments from panels (a) to (d). It showcases the localization patterns of FMR modes under distinct magnetic field configurations, represented within $4\times4$ column arrays for detailed visual comparison.
\label{Fig:ladd_loc}}
\end{figure}

The scaffold structure is subjected to micromagnetic simulations similar to those performed on gyroids. We examine the FMR response of the structure as a function of the 500~mT external magnetic field direction within its $xy$-plane. Representative FMR spectra for the two analyzed scaffold structures in the field at 0 and 45~deg (from $x$-axis, in-plane) are shown in Fig.~\ref{Fig:scaff_fmr}(a) and (b), respectively. They were determined using the same technique as described above for the gyroids (Fig.~\ref{Fig:gyr_fmr}) and explained in detail in Methods, Sec.~\ref{Sec:sim}. Again, the most intense mode has the lowest frequency and oscillates in phase throughout the volume. There is also the second intense peak clearly visible in Fig.~\ref{Fig:scaff_fmr}(a) for a static field angle of 0~deg (blue line, 9.80~GHz) which represents the first, asymmetric SW mode quantized along the $z$-direction [see Fig.~S1(c) and (d) in the Supplementary Information].

Initiating with the field oriented along the $x$-axis [Fig.~\ref{Fig:ladd_loc}(a)], the symmetric configuration exhibits bulk localization (frequency 9.79~GHz), namely in nanorods oriented perpendicular to the external field (these include $y$-axis aligned rods and connecting vertical bars) with almost no intensity in the nanorods aligned with the magnetic field. Conversely, the asymmetric configuration demonstrates energy localization within the upper plane of the film, resulting in a surface mode (9.59~GHz). A 45~deg rotation of the field, [Fig.~\ref{Fig:ladd_loc}(b)], yields a configuration where the nanorod junctions became focal points for SW amplitude concentration, favoring the bulk section. The orientation of the field along the $y$-axis [Fig.~\ref{Fig:ladd_loc}(c), 90~deg] brought forth a critical scenario in this simulation segment. For the symmetric structure, where the nanorods aligned with the $x$-axis are present, localization at both surfaces is observed (at 9.58~GHz). In contrast, in the asymmetric structure, the localization is manifested on the bottom surface, i.e., opposite to the $x$-axis saturation case at 9.59~GHz. Further on, a 135~deg rotation [Fig.~\ref{Fig:ladd_loc}(d)] repeats the results obtained for 45~deg rotation. The rotation of 180~deg brings back the original structure, completing the cycle.

\subsection{SW localization -- quantitative analysis}

To accurately quantify SW localization, the inverse participation ratio (IPR) serves as a valuable parameter. Traditionally employed in quantum mechanics to assess the localization of wave function~\cite{Calixto_2015, He_2022, Berciu_2002, Szallas_2008} the IPR is defined as $\text{IPR} = \sum_i |\psi_i|^4 / (\sum_i |\psi_i|^2)^2$, where $\psi_i$ symbolizes the wave function at the $i$-th site or lattice point. This parameter effectively measures how concentrated the wave function is within a given discrete space, providing a scalar value that differentiates localized states from extended ones.
In continuous ferromagnetic systems, as in our case, the IPR needs the transition to the continuous form, which requires an integral form~\cite{Imagawa_2010} and the use of a SW amplitude instead of the wave function. Finally, the formula for IPR in magnonic systems can be expressed as:
\begin{equation}
    \text{IPR} = \frac{\int_V |m(\mathbf{r})|^4 dV}{(\int_V |m(\mathbf{r})|^2 dV)^2},
    \label{Eq:ipr_m}
\end{equation}
where $|m(\textbf{r})| =(m_x(\textbf{r})m_x^\ast(\textbf{r})+m_y(\textbf{r})m_y^\ast(\textbf{r})+m_z(\textbf{r})m_z^\ast(\textbf{r}))^{1/2}$ is the position-dependent absolute value of the complex SW amplitude, and the integration is over the volume $V$ of the single UC. An asterisk sign (*) indicates a complex conjugate. For a completely delocalized magnetic excitation, where the amplitude of the SW mode is uniform across the entire volume of the ferromagnet, the IPR yields the value of 1. Conversely, in the case of extreme localization, where the SW mode is concentrated at a single point within the volume, resembling the behavior of a Dirac delta function, the IPR approaches infinity.

\begin{figure}[htp]
\includegraphics[width=\linewidth]{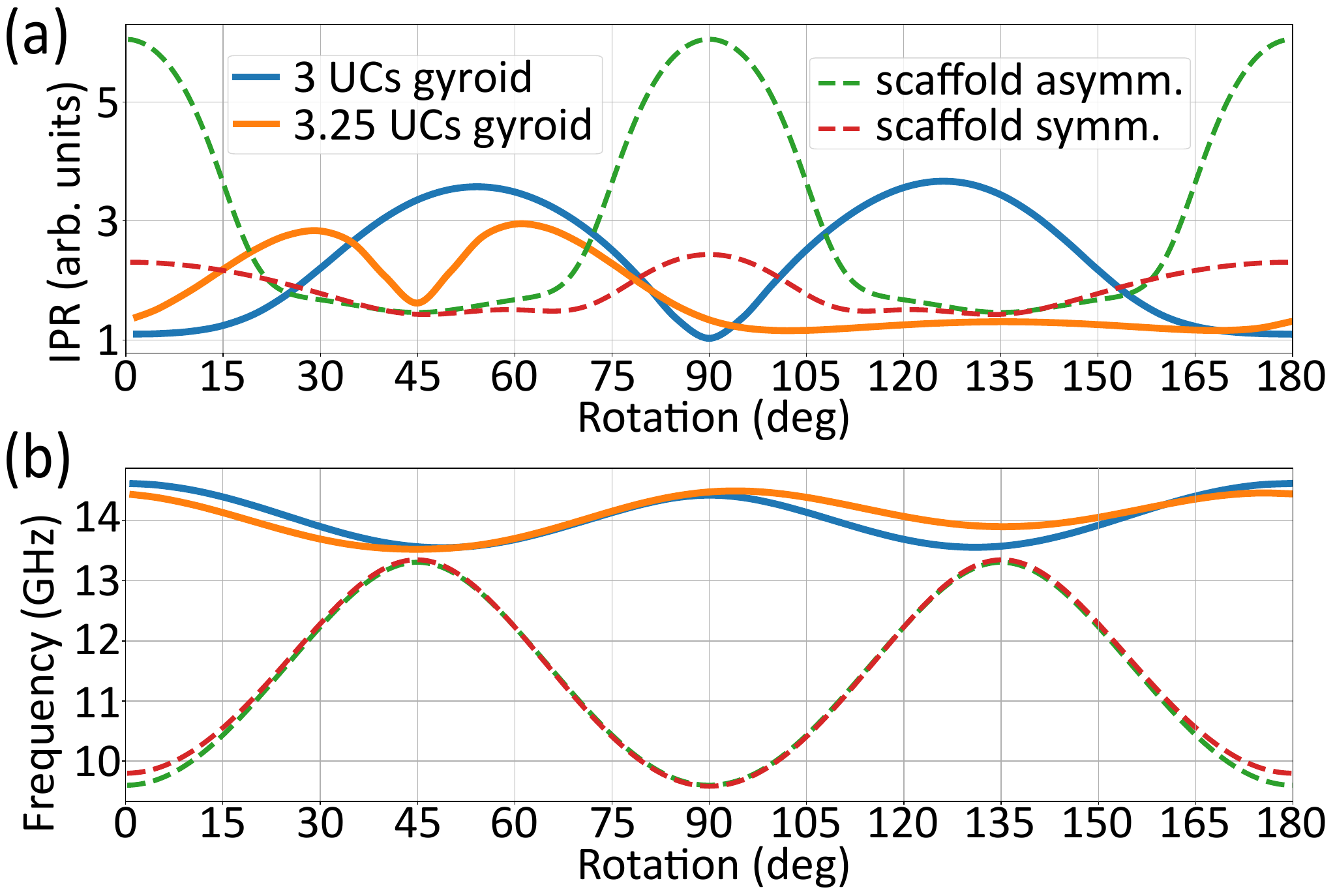}
\caption{IPR (a) and FMR frequency (b) as a function of the rotation angle of the external magnetic field over the plane of the analyzed gyroids (solid line) and scaffolds (dashed line) structures. Different colors indicate the asymmetric and symmetric configurations of both nanosystems.
\label{Fig:all_ipr}}
\end{figure}

Fig.~\ref{Fig:all_ipr}(a) provides a quantitative validation of the SW mode localization behavior inferred from the results depicted in Figs.~\ref{Fig:gyr_loc} and \ref{Fig:ladd_loc}. The IPRs of the scaffold structures (dashed lines) show that the localization of SWs occurs at field angles of 0, 90, and 180~deg. From Fig.~\ref{Fig:ladd_loc}, we can see that this corresponds to the top-bottom-top cycle of surface localization in asymmetric scaffolds, and the center-surfaces-center magnetization distribution in symmetric scaffolds, respectively. The lower IPR values for the symmetric structure (red dashed line) are due to the fact that the SW modes are generally distributed over a larger volume than in the asymmetric case (green dashed line). This result further underscores the conclusions of Sec.~\ref{sec:ladd}, i.e., the crucial influence of structural symmetry on the magnetization switching within the scaffolds, making the localization on opposite surfaces energetically preferential for the asymmetric structure after a 90-deg field rotation.

The angular dependence of IPR for 3-UCs gyroids [blue solid line in Fig.~\ref{Fig:all_ipr}(a)] qualitatively mirrors (with a 45~deg shift) the behavior seen in the asymmetric scaffold structure. However, an IPR maximum is observed at field rotation angles around 55~deg and, due to symmetry, around 125~deg. Conversely, the IPR values decrease at angles of 0, 90, and 180~deg. In the case of 3.25-UCs gyroids (orange solid line), the situation changes significantly -- here we have a clear weakening of the IPR (due to both side localization) and a shift of the IPR peaks to the first quarter of the bias field angle, and a flattening of the IPR to a low value between 90 and 180~deg of the bias field orientation. Thus, the two gyroidal structures, which differ in height by only 0.25~UC, show significant differences in these dependencies. This indicates the effect of the surface cut and the breaking of the 90-deg symmetry, suggesting the influence of structural chirality.

The strong influence of the surface cut on the localization phenomenon can be understood by looking carefully at the cuts in Fig.~\ref{Fig:gyr_loc} -- the top and bottom surface struts have their specific effective directions relative to the external magnetic field. However, the part primarily responsible for the localization is not the outermost surface, but its inner junction towards the center [can be seen well in the upper left corner image in Fig.~\ref{Fig:gyr_loc}, and the first illustration in Fig.~\ref{Fig:ladd_str}(c)], as indicated by the slightly higher SW amplitude. This is due to a larger demagnetizing field as will be discussed later in this paper. Thus, in a system composed of 3~UCs, the line effectively normal to this part of the structure is directed about 125~deg from the $x$-axis in the upper layer and about 55~deg in the lower layer. For a 3.25~UC high gyroid, the normal to the SW localization inducing part is rotated about 35 and 55~deg from the $x$-axis for the top and bottom layers, respectively. As can be seen from the plots in Fig.~\ref{Fig:all_ipr}(a), this coincides well with the IPR peaks. We can therefore correlate the surface cut-localization dependence with the demagnetizing field and chirality of the gyroidal structure. It is also worth noting that the effect of SW localization is always present in gyroids -- it is only its dependence on the direction of the bias magnetic field that changes (more examples in the Supplementary Information, Sec.~II, Figs.~S4 and S5).

In Fig.~\ref{Fig:all_ipr}(b) we see the resonant frequency of SW cyclically decreasing and increasing with the field rotation, which is clearly associated with the localization peaks in all cases. The maximum frequency variation is approximately 3.5~GHz for scaffolds and 1~GHz for gyroids. There is a one-to-one correspondence between the maximum (minimum) of the IPR and the minimum (maximum) of the FMR frequency for the scaffold structure. Consequently, the frequency dependence is a smooth trigonometric function characteristic of the two-axis easy anisotropy system. However, a slight asymmetry is observed for the gyroid (solid orange line) between the orthogonal directions of the magnetic field, again indicating that the chirality of the structure may play a role. Thus, the fourfold symmetry seen in the FMR spectra is also present in the IPR dependence on field orientation for the scaffold structures, but it is lost in gyroids. The drop in the mode frequency with increasing IPR can be understood as an influence of the static demagnetizing field resulting in the accumulation of dynamic magnetization in smaller regions. 

\subsection{Demagnetizing field effect}
Common to woodpile-like systems is that the lowest frequency mode is always localized in the nanorods oriented perpendicular to the external magnetic field. This is a result of the static demagnetizing field. To better understand this effect, we take the smallest building block of the scaffold structure, which is a single infinitely long nanorod~\cite{Chen1991}. When the field is parallel to it, the demagnetizing field is not generated, resulting in a high FMR frequency of 22.41~GHz~\cite{kittel}. On the other hand, the external field perpendicular to the rod produces a strong demagnetizing field, resulting in a reduction of the effective magnetic field and lowering the FMR frequency of a single rod to 8.94~GHz. The field rotated by 45~deg to the rod produces a moderate demagnetizing field so that the FMR frequency reaches 16.00~GHz. This behavior is effectively transferred to the scaffold structure, shown in Fig.~\ref{Fig:ladd_loc}. For 0 and 90~deg, the FMR mode is localized in the levels with rods perpendicular to the external field (9.79 and 9.58~GHz). For 45 and 135~deg configurations (13.35~GHz), only the vertical bars remain perpendicular to the field, and therefore they are the parts with the strongest amplitude of the FMR mode. 

\begin{figure}[htp]
\includegraphics[width=\linewidth]{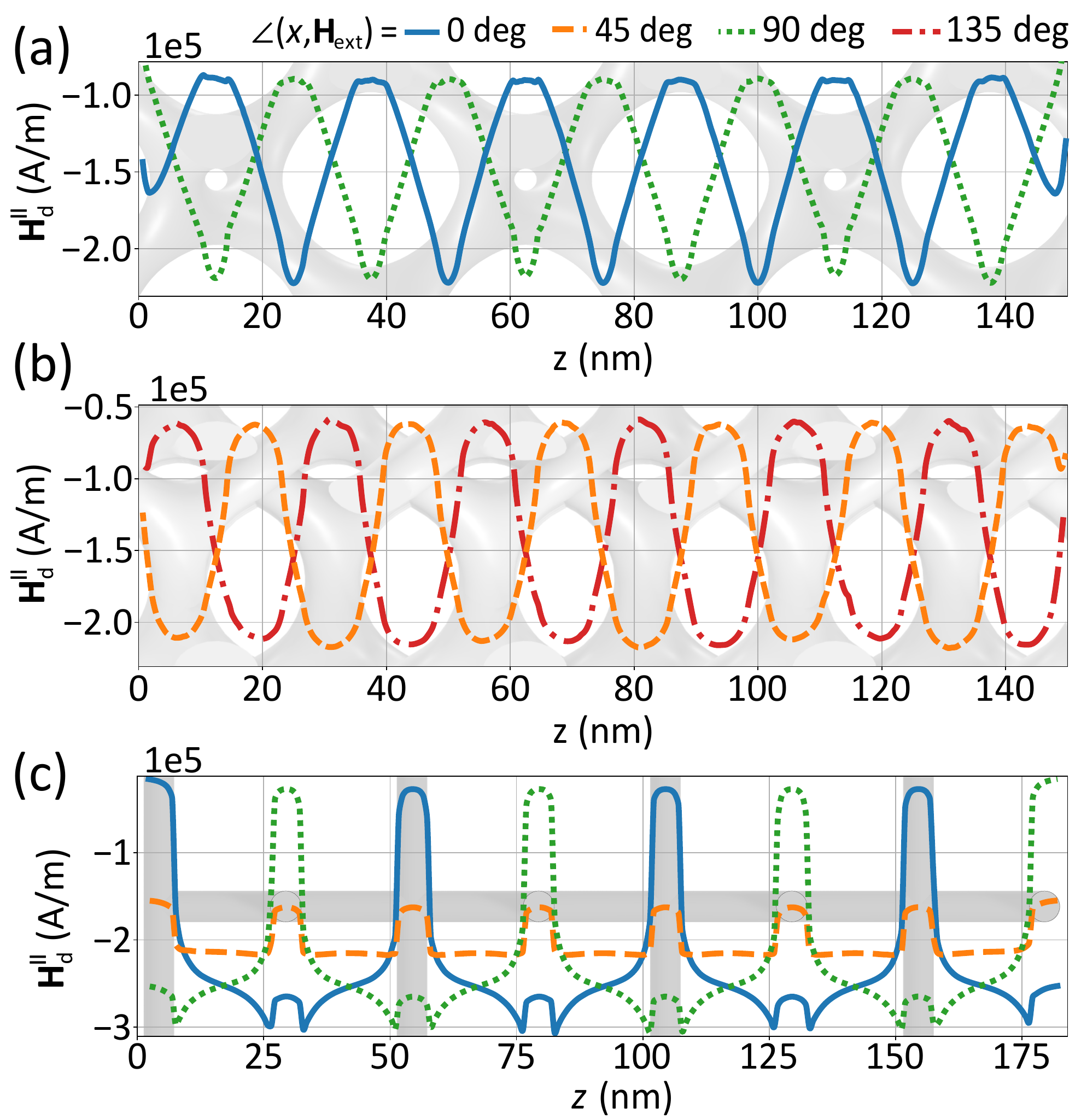}
\caption{Plots of the internal magnetostatic field $\textbf{H}_\text{d}^{\parallel}$ (including demagnetizing and stray fields, see Methods~\ref{Sec:sim}) parallel to the bias field $\textbf{H}_\text{ext}$, averaged over the UC and projected here on the $z$-axis (along its height, $\textbf{H}_\text{d}^{\parallel}(z)=1/S\int_{S}\textbf{H}_\text{d}(x,y,z)dxdy$). The different colors correspond to the average distribution of $\textbf{H}_\text{d}^{\parallel}$ for different directions of the static magnetic field $\textbf{H}_\text{ext}$ relative to the $x$-axis. Panel (a) shows the demagnetizing field in the 3-UCs gyroid for the $\textbf{H}_\text{ext}$ field directed at 0 and 90~deg (the corresponding projection of the structure can be seen in the background). In (b) we see the analogous results for angles of 45 and 135~deg, along with the corresponding projection of the gyroid. (c) The $\textbf{H}_\text{d}^{\parallel}$ of the asymmetric scaffold structure, for an external magnetic field directed at angles of 0, 45 and 90~deg (background structure projection for normal along the $x$-axis).
\label{Fig:gyr_vs_ladd}}
\end{figure}

As shown in Fig.~\ref{Fig:ladd_str}(c), both gyroids and scaffold structures are built from a vertically alternating distribution of bonds. This leads to an analogous dependence of the static demagnetizing field on the external field rotation as for the single nanorod described above. If the field is effectively parallel to the gyroid struts (and at the same time perpendicular for adjacent ones), the local demagnetizing field will be alternately weaker and stronger, creating the potential wells in which the FMR modes are localized. Although perfect alignment is impossible in gyroids due to the lack of straight rods (see Fig.~\ref{Fig:gyr_str}), we observe analogous properties of the magnetization distribution as in the scaffold nanostructures.

When discussing the demagnetizing field, we cannot overlook the stray field generated by the nanorods. As an example, let us take the symmetric scaffold structure for the field applied along the $x$-axis, as depicted in Fig.~\ref{Fig:ladd_loc}(a). In this case, the rods aligned with the $y$-axis (with the largest mode amplitude) produce a stray field in the neighboring levels of parallel nanorods, in the direction opposite to the external field. Levels in the center of the structure have two close parallel neighbors compared to one for the surface level, so the stray field is stronger in the center. Since the stray field enhances the effect of the demagnetizing field, the effective field in the center is the smallest, resulting in the lower local FMR frequency and, hence, the largest amplitude. The same reasoning can be applied to the 45 and 135~deg cases. However, the stray field cannot explain the strong surface localization that is present for the 0-deg field in the asymmetric structure and for the 90-deg field in both structures. This is further confirmed in Fig.~\ref{Fig:gyr_vs_ladd}(c), which shows the UC-averaged internal magnetostatic field $\textbf{H}_\text{d}^{\parallel}$ (including the demagnetization as well as a stray field) along the $z$-axis for the asymmetric scaffold structure for 0, 45, and 90~deg field orientation. It is clear that the depths in the magnetostatic field are located at the nanorods, which are perpendicular to the bias field at 0 and 90~deg, but their depth is almost the same. Qualitatively similar dependencies of the average magnetostatic field are found in the gyroid [Fig.~\ref{Fig:gyr_vs_ladd}(a,b)]. However, here the field variations are smooth due to the complex geometry and different orientations of the struts. Thus, the inhomogeneity of the internal magnetostatic field determines the type of struts of the FMR mode localization, but does not justify the surface or bulk localization of the SW mode in dependence on the bias field orientation. For more information on the effect of the gyroidal filling factor and its height on localization, see the Supplementary Information, Sec.~II.

The results for both scaffold and gyroid structures not only underscore the intricate interplay between geometry, magnetic field orientation, and magnetostatic fields in these advanced magnonic materials but also point to the important role of other interactions in enabling the tunable localization of SW modes. The mode analysis during the design of a woodpile-like structure suitable to reproduce this effect (Supplementary Information, Fig.~S6) showed the necessity of its vertical continuity, which led us to conclude that the surface localization in 3D nanostructures is not related to the dynamic stray field, but rather to the exchange interaction. The following section and simulations confirm that exchange-related effects are critical in driving this behavior.

\subsection{Exchange interaction and vertical period impact in scaffold structure \label{Sec:d_exch}}

\begin{figure*}[htp]
\includegraphics[width=\linewidth]{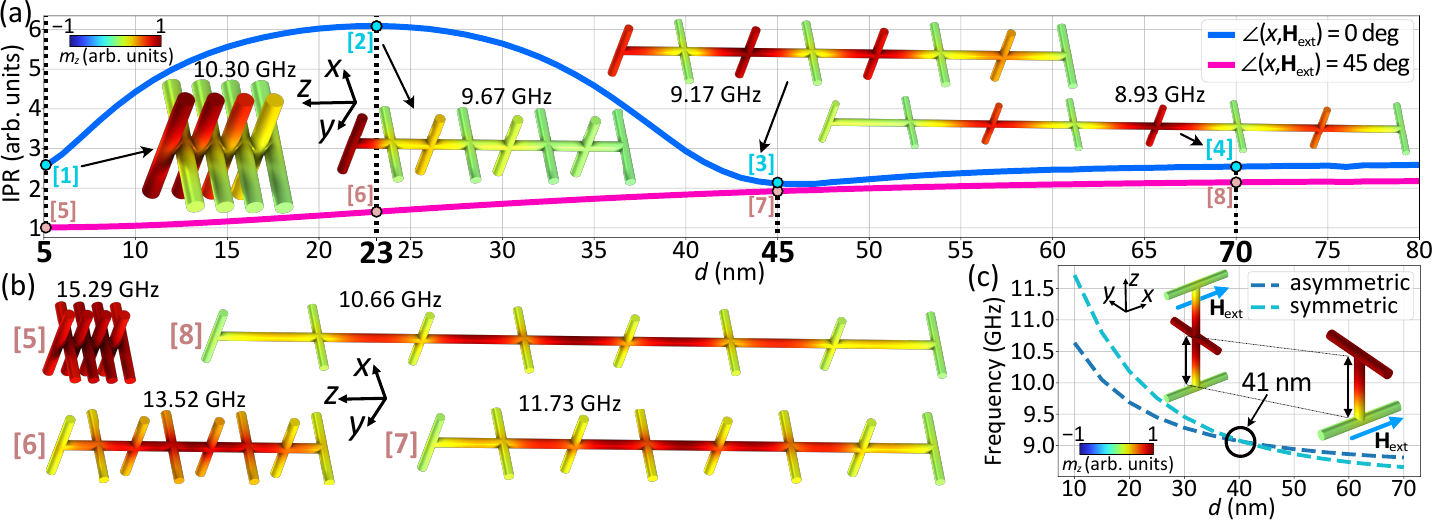}
\caption{The ferromagnetic response of the scaffold nanosystem as a function of the distance ($d$) between adjacent nanorods. Panel (a) presents the IPR against $d$ under two conditions: with the magnetic field aligned along the $x$-axis (depicted in blue) and with the field rotated by 45~deg within the plane (shown in magenta). The critical points highlighted on this curve correspond to specific distributions of the dynamic magnetization component $m_z$ of the FMR modes within the analyzed nanostructures. These key spots are marked by numbers, with visualizations of the modes for the 45-deg field configuration provided in (b). Panel (c) illustrates how the FMR frequency varies with $d$ for both symmetric and asymmetric one-unit cell high scaffolds, for $\textbf{H}_\text{ext}$ parallel to the $x$-axis. It includes depictions of the modes at a significant juncture -- where the frequencies of both configurations converge at $d=41$~nm.
\label{Fig:ipr_vs_d}}
\end{figure*}

Micromagnetic simulations investigating the variation of the vertical distance between neighboring nanorods, $d$, in asymmetric scaffold structures, as shown in Fig.~\ref{Fig:ipr_vs_d}, support our interpretation of the influence of exchange interactions on the observed effects. This analysis focuses on the amplitude distribution of FMR mode and corresponding IPR values as a function of the separation between adjacent perpendicular nanorods. Fig.~\ref{Fig:ipr_vs_d}(a) shows a pronounced IPR value for 0~deg (blue line) within the 5-45~nm range of $d$, with the maximum $\text{IPR}\approx6$ at 23~nm, signifying the FMR mode localization on the surface nanorod (see the modes $\langle1\rangle$ and $\langle2\rangle$). Beyond this optimal distance, the magnetization shifts from a surface localized state to the bulk (mode $\langle3\rangle$), saturating IPR value at about 2.5. The bulk mode concentration remains in nanorods, which are perpendicular to the field, reflecting the inherent inability of the structure to adopt a fully delocalized configuration across its volume (mode $\langle4\rangle$). For simulations with the field oriented 45~deg from the $x$-axis [indicated by the magenta curve in Fig.~\ref{Fig:ipr_vs_d}(a)], the response is monotonic, increasing IPR from 1 at $d=5$~nm up to 2.1 at $d=80$~nm. This change in IPR is associated with the change in the SW amplitude distribution from uniform at $d=5$~nm (mode $\langle5\rangle$) to the bulk with amplitude in the nanorod oriented along the $z$-axis (mode $\langle8\rangle$).

In both cases, 0 and 45~deg, the amplitude distribution at $d>45$~nm corresponds to the $\mathbf{H}_\text{d}^{\parallel}$ profiles [see Fig.~\ref{Fig:gyr_vs_ladd}(c)], i.e., the maximum of the SW amplitude is concentrated in the regions with the largest $\mathbf{H}_\text{d}^{\parallel}$. The proximity between $(x,y)$ plane-oriented nanorods at $d<45$~nm narrows these potential wells for SW confinement, increasing their frequency and the leakage of the amplitude into neighboring regions, especially in the case of shallow wells of $\mathbf{H}_\text{d}^{\parallel}$, i.e., at 45~deg where well depth is less than 50~kA/m [Fig.~\ref{Fig:gyr_vs_ladd}(c)]. In the case of 0~deg, the potential wells are deep (above 250~kA/m), and the increase in frequency is associated with the transfer of the SW amplitude to the surface, which has a nanorod perpendicular to $\mathbf{H}_\text{ext}$, since it provides more suitable conditions for the FMR mode than the bulk cells. This is because in a bulk part, each field-orthogonal rod is flanked by two field-parallel "pinning" neighbors, whereas surface nanorods are influenced by only one such neighbor. Thus, this phenomenon can be attributed to the exchange interactions that make the frequency of SWs and their localization dependent on the magnetization pinning at the boundaries: the demagnetizing field wells or surfaces. Consequently, at 0~deg and $d<45$~nm the only one-sided pinning of the SW amplitude in the surface field-orthogonal nanorod provides the suitable conditions for lowering the frequency of the surface-localized SW [modes $\langle 1 \rangle$ and $\langle 2 \rangle$ in Fig.~\ref{Fig:ipr_vs_d}(a)]. This effect is similar to SW surface localization and SW quantization in thin ferromagnetic films~\cite{Yu1975,Puszkarski1979,Levy1981}. However, here the pinning/unpinning is introduced by surface anisotropy at the surface of the atomic lattice of spins, resulting in bulk/surface SW formation.

\begin{figure}[htp]
\includegraphics[width=\linewidth]{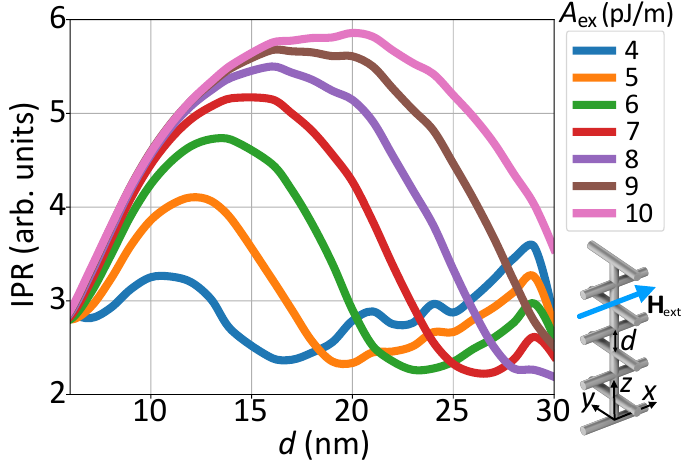}
\caption{Plot of IPR variation within the range indicative of surface localization in scaffold structures (asymmetric, $\textbf{H}_\text{ext}$ along $x$-axis), as a function of the nanorod separation $d$, for different values of the exchange constant $A_\text{ex}$.
\label{Fig:A_vs_d_vs_IPR}}
\end{figure}

Fig.~\ref{Fig:ipr_vs_d}(c) illustrates how increasing $d$ influences FMR mode frequencies in both symmetric and asymmetric single-unit scaffold cells for a bias magnetic field parallel to the $x$-axis (0~deg). The analysis of a single field-perpendicular rod with one magnetization-fixing rod oriented parallel to the field for the asymmetric case, and two for the symmetric case, excludes the influence of stray magnetostatic field interactions from neighboring field-perpendicular nanorods. This allows an isolated analysis of the effect of the proximity of the adjacent rods. In both configurations, the FMR mode is primarily concentrated in the nanorod oriented perpendicular to the field. In the symmetric structure, it occurs in the bulk, while in the asymmetric structure, it is localized at the surface. Notably, the frequencies of the FMR in both configurations converge at $d=41$~nm, with the lowest frequency for surface localization at $d<41$~nm and bulk at $d>41$~nm. This suggests that the spatial separation along the $z$-axis between rods reaches a threshold at $d=41$~nm, beyond which exchange energy no longer dominates the dynamic magnetization distribution. This observation aligns with Fig.~\ref{Fig:ipr_vs_d}(a) for 0~deg field orientation, where the scaffold structure's localization similarly diminishes around 45~nm, marking the shift of localization from surface to bulk rods, with minor discrepancy in $d$ value.

To further confirm that the surface localization in scaffolds within a small $d$ range is an exchange interaction effect, we conducted additional simulations varying the exchange constant ($A_{\text{ex}}$), and plotting IPR($d$). The results for asymmetric scaffold structure at 0~deg are depicted in Fig.~\ref{Fig:A_vs_d_vs_IPR}. There is a pronounced variation in the IPR values, illustrating ultimately the significant impact of exchange interactions on the localization phenomena. Specifically, at $A_{\text{ex}}=4$~pJ/m, the IPR profile appears weaker and irregular, indicating subdued surface localization effects. In contrast, at $A_{\text{ex}}=10$~pJ/m, there emerges a distinct range ($d<35$~nm), where surface localization is enhanced [max(IPR) $\approx 6$], demonstrating a clear and strong SW localization effect. The exchange dominance effects on the surface localization of the FMR mode shown above for the scaffold structure can be related to the gyroid structure and the surface localization observed in Fig.~\ref{Fig:gyr_loc}, but due to different structures and smooth demagnetizing field variation, it needs additional analysis.

\subsection{Effect of exchange interaction on surface localization in gyroids \label{sec:gyr_exch}}

In scaffolds, we can directly manipulate the structural parameter ($d$), and its relation to $A_\text{ex}$. In gyroids, we can use the filling factor $\phi$ (see the Supplementary Information, Sec.~II.A), but by changing it we collectively affect the entire geometry and the nature of all interactions in it, including magnetostatics, shape anisotropy, and the ratio of the gyroid linear dimensions to the exchange length. In particular, the flattening of the demagnetizing field with increasing $\phi$ correlates with decreasing IPR and decreasing modulation of the FMR frequency (see Fig.~S2 in the Supplementary Information). Thus, in this section, we analyze the gyroid structure with 3~UCs and keep the filling factor at $\phi=10\%$ (parameters as in Sec.~\ref{sec:h_rot}).

\begin{figure}[htp]
\includegraphics[width=\linewidth]{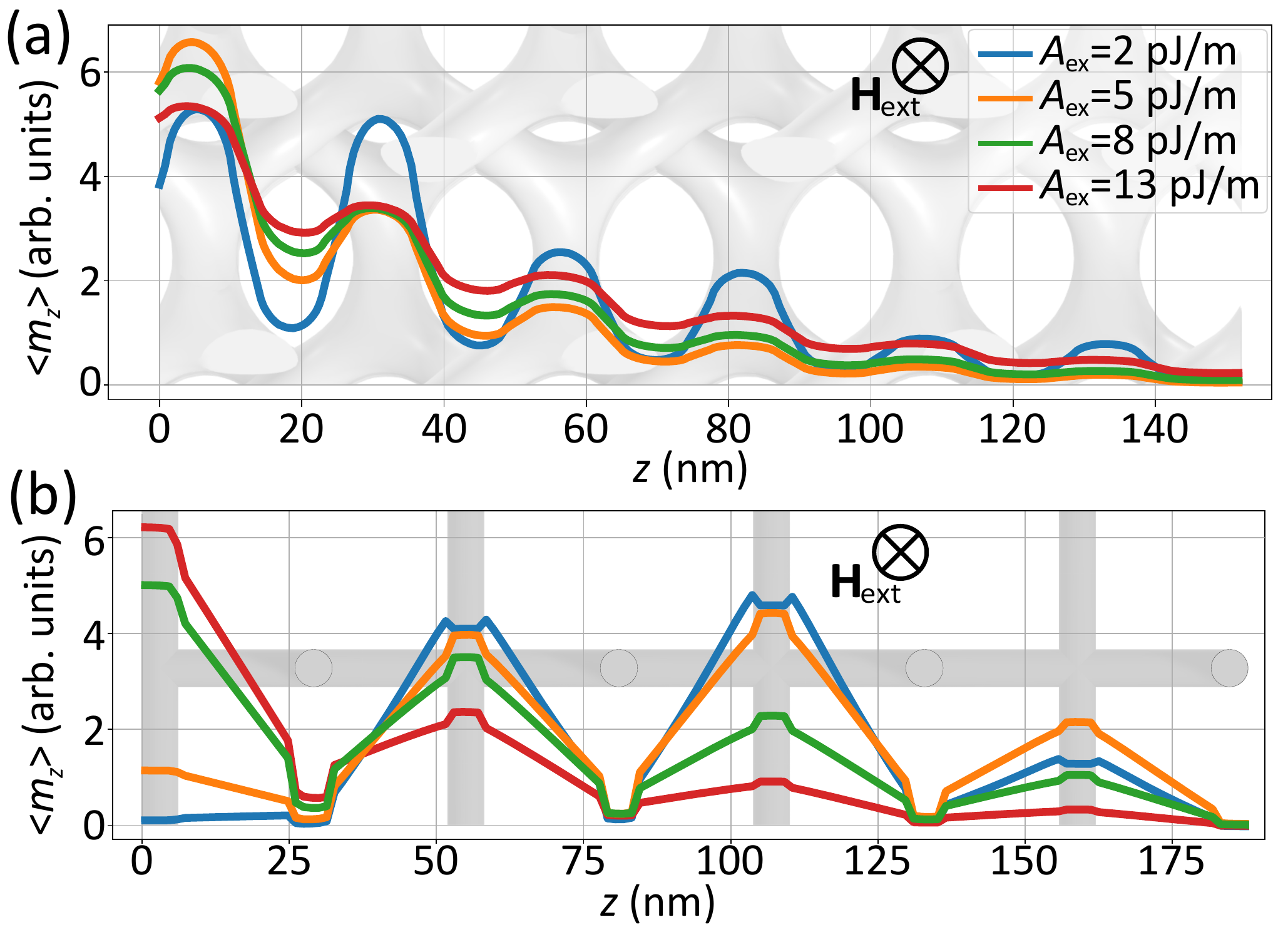}
\caption{Plots of the distribution of the dynamic magnetization component $m_z$, averaged and projected on the $z$-axis, for the gyroid structure with 3~UCs (a) and the asymmetric scaffold structure (b). For comparison, the case with the field at an angle of 45~deg from the $x$-axis was used for the gyroid, and for scaffold along the $x$-axis (0~deg) -- both cases show clear surface localization at large $A_\text{ex}$ (see Figs.~\ref{Fig:gyr_loc} and \ref{Fig:ladd_loc}). The colors represent different values of the exchange constant $A_\text{ex}$, whose legend is common to both plots.
\label{Fig:projM_vs_A}}
\end{figure}

In Fig.~\ref{Fig:projM_vs_A}, the localization effect is shown in the form of one-dimensional projections of the $z$-component of the dynamic magnetization along the $z$-axis, averaged over the UC in the $(x,y)$ plane. The results of the gyroid [Fig.~\ref{Fig:projM_vs_A}(a)] are juxtaposed with those of the scaffold [Fig.~\ref{Fig:projM_vs_A}(b)] to highlight the differences and similarities of the exchange contribution to the localization effect of the studied SW modes. 

In Fig.~\ref{Fig:projM_vs_A}(b) we see the magnetization curves $\langle m_z\rangle$ for an asymmetric scaffold structure with neighboring rods at $d=25$~nm. There is a strong dependence of the surface localization on the exchange constant -- the transition from the bulk mode for small $A_\text{ex}$ ($\le 5$~pJ/m) to the surface mode for $A_\text{ex}\ge 8$~pJ/m. We see a similar, though smaller effect of $A_\text{ex}$ on the surface localization of the FMR mode in the gyroid. For $A_\text{ex}=2$~pJ/m the SW amplitude is almost equal in the first two perpendicular rods, i.e., a clear shift of the dynamic magnetization towards the center of the layer can be observed [blue line in Fig.~\ref{Fig:projM_vs_A}(a)]. For larger values of the exchange constant, the surface localization is preserved, and the smoothing of the $\langle m_z\rangle$ curves along the $z$-direction with increasing $A_\text{ex}$ is observed. A larger exchange ensures that the magnetization is not only concentrated in bars perpendicular to the field (where the demagnetization is largest), but spreads more homogeneously to neighboring struts, analogous to scaffolding structures (for sufficiently small $d$ and large exchange length). A more pronounced effect of the transition from the bulk to the surface state as a function of the exchange constant occurs for a gyroid with a filling of $\phi=20\%$, for which results can be found in Fig.~S3 in the Supplementary Information. In addition, we performed an analysis of the effect of the height of the gyroidal layers (independent of the cut point) on SW localization, which further confirms the influence of exchange energy on the presence of surface localization. The results are shown in the Supplementary Information, Figs.~S4 and S5.

Based on the analysis of the localization of SWs as a function of the exchange constant in the studied gyroid structures and their comparison with the scaffolds along the $z$-direction, we can conclude that it has a different, though fundamental, influence in both cases. In gyroids, the transition from the bulk to the surface state is determined only for very small values of $A_\text{ex}$ (a relationship strongly related to the filling factor). A stronger exchange determines the uniformity of magnetization within neighboring nanowires. Woodpile-like scaffolds show a more "stepped" and monotonic $A_\text{ex}$-related transition from bulk to surface localization of SWs.

\section{Discussion}

In this study, we explored surface localization of the FMR mode phenomena within thin films made of gyroid and woodpile-like scaffold three-dimensional ferromagnetic nanostructures, focusing on the effects of the in-plane external magnetic field rotation. Using micromagnetic simulations, we have demonstrated a novel surface localization of SWs that differs from other known wave localization phenomena. Unlike Damon-Eshbach localization, which requires SW propagation, or Shockley and Tamm surface states and topologically protected edge modes, which rely on a Bragg bandgap, this newly observed surface localization does not satisfy these requirements. It also differs from edge-localized magnetostatic modes that occur in the demagnetization wells oriented perpendicular to the surface. Instead, we found that the surface localization in considered 3D structures is a cooperative effect of the magnetostatic (demagnetizing and stray) in-plane field and exchange interactions. The former creates potential wells in the nanorods (gyroid struts) perpendicular to the bias magnetic field, the latter determines the frequency in the well when its width is comparable to the exchange length. As a result, for some bias field orientations and surface cuts the SW localized at the surface UC has only one-sided pinning, which lowers its energy, making it a low-frequency surface-localized FMR mode. Thus, this research highlighted the critical influence of the static demagnetizing fields and the exchange energy in shaping the SW amplitude distribution in the ferromagnetic response of such nanostructures, thereby contributing to our understanding of the magnonic behavior in 3D structures.
Furthermore, the intricate relationship between the magnetic field orientation and the geometry of the structures was revealed, i.e., the surface configuration seemed to strongly influence the SW amplitude concentration along the height of the thin film. Nevertheless, the localization for a given field direction persists over different surface states, demonstrating its universal nature.

Such selective localization of the FMR mode introduces a novel mechanism enabling reconfigurable functionalities. This reveals the potential for enhancing experimental measurements of SWs in three-dimensional structures through localized FMR modes, among others, in the established optical techniques like Brillouin light scattering (BLS)~\cite{Sebastian2015} or magneto-optical Kerr effect (MOKE) microscopy~\cite{Perzlmaier2008}. This advancement promises improved sensitivity in probing SW dynamics, offering deeper insights into the magnetic properties of complex structures. The findings gained from this work may also open new avenues for device design that exploit the unique properties of gyroids, woodpile-like scaffolds, or other 3D nanoarchitectures to advance the development of next-generation magnonic technologies in 3D.

\section{Methods}

\subsection{Micromagnetic simulations \label{Sec:sim}}
To calculate SW modes within the 3D nanostructures, we employed the COMSOL Multiphysics software. It harnesses the finite-element method (FEM) to provide solutions to complex coupled systems of partial differential equations. 
The SW dynamic is framed with the Landau-Lifshitz (LL) equation:
\begin{equation}
\frac{\partial \mathbf{M}}{\partial t} = -\gamma\mu_0\mathbf{M} \times \mathbf{H}_{\text{eff}}
\label{Eq:LL}
\end{equation}
where $\mathbf{M}$ is the magnetization vector, $\gamma$ denotes the gyromagnetic ratio, $\mu_0$ is the vacuum permeability, and $\mathbf{H}_{\text{eff}}$ is the effective magnetic field. The nonuniformity in material properties (e.g., variations in $M_\text{s}$, magnetic anisotropy, or exchange stiffness) is one of the ways that introduces a spatially dependent $\mathbf{H}_{\text{eff}}$ that, in turn, affects the localization and dispersion properties of SWs. Here, it merges the externally applied field, $\textbf{H}_\mathrm{ext}$, with the magnetostatic demagnetizing field, $\textbf{H}_\mathrm{d}$, and the Heisenberg exchange field, $\textbf{H}_\mathrm{exch}$:
\begin{equation}
\textbf{H}_\text{eff}=\textbf{H}_\text{ext}+\textbf{H}_\text{d}+\textbf{H}_\text{exch}.
\label{eq:heff}
\end{equation}
The demagnetizing field is critical for SW dynamics in ferromagnetic materials, especially when it is patterned. Governed by Amp{\`e}re's law, this field is derived from the gradient of the magnetic scalar potential, $\textbf{H}_\text{d}=-\nabla U_\text{m}$. Within the magnetic body, this relationship further evolves in:
\begin{equation}
\nabla^2 U_\text{m}=\nabla\cdot\textbf{M},
\label{Eq:nab2}
\end{equation}
while outside it, $\nabla^2 U_\text{m}=0$. In performed COMSOL implementation, we tackled the eigenproblem derived from Eqs.~(\ref{Eq:LL}),~(\ref{eq:heff}), and~(\ref{Eq:nab2}). By presuming full magnetization saturation via the bias magnetic field and adopting a linear approximation, we could dissect the magnetization vector into its static and dynamic (time $t$ and position $\textbf{r}$ dependent) components $\textbf{M}(\textbf{r},t) = M_\text{s} \hat{i} + \delta \textbf{M}(\textbf{r},t)\;\forall\;(\delta \textbf{M}\perp\hat{i})$, neglecting all nonlinear terms in the dynamic magnetization $\delta\textbf{M}(\textbf{r},t)$. Here, we assume that the static component of the magnetization is equal to the saturation magnetization, $M_\text{s}$. This methodology is further explained in Refs.~\cite{Mruczkiewicz2013StandingCrystals,Rychy2018SpinRegime}.

Using PBCs on the UC boundaries along the $x$- and $y$-axes, we model in COMSOL an infinite in-plane gyroidal and scaffold-structured films (see Figs.~\ref{Fig:gyr_str} and \ref{Fig:ladd_str}). The PBCs are defined on both faces to maintain the same values for the magnetization components and magnetic scalar potential. For planes parallel to the surfaces of the films, we implemented Dirichlet boundary conditions aiming to suppress the scalar magnetic potential, $U_\text{m}=0$, at the boundaries of the computational cell. To ensure the simulation's physical accuracy and convergence, it is essential to position these conditions sufficiently far from the specimen. In our simulations, the computational cell's height was set to be 40 times the gyroid/scaffold layer's height.

Throughout the simulations, consistent mesh quality was maintained across the different gyroid and scaffold models. The quality of the tetrahedral discretization mesh, characterized by the volume-to-length parameter, remained stable at an average value of about 0.7. It is based on a ratio of element edge lengths to element volume. This resulted in a scalable mesh of about 55000 elements for the single cubic $\phi=10\%$-gyroid model and about 22000 for a 50~nm long nanorod.

Calculations of the FMR spectra (Figs.~\ref{Fig:gyr_fmr} and~\ref{Fig:scaff_fmr}) of the studied structures were obtained by simulations in the frequency domain, sweeping the spatially uniform, dynamic microwave field in a given range with a step of 50~MHz. Its magnitude was set to $\mu_0 h_\text{dyn}=0.005\mu_0 H_0=2.5$~mT and was polarized along the $y$-axis. To determine the macroscopic measure of the global magnetization intensity, the complex dynamic component $m_z$ (perpendicular to both the static and dynamic fields) was multiplied by its conjugated value $m_z^{*}$, and integrated over the entire volume of the ferromagnet:
\begin{equation}
    I=\int_V m_z m_z^{*} dV.
    \label{eq:int}
\end{equation}
The plots for each structure have been normalized to the maximum of one of the two spectra (the one with higher maximum intensity), preserving their relative ratio.

\section*{Data Availability}
The data underlying this study are openly available in Zenodo
at \url{https://doi.org/10.5281/zenodo.13141840}.

\bibliography{main}

\begin{acknowledgments}
The research leading to these results was funded by the National Science Centre of Poland, Projects No.~UMO-2020/39/I/ST3/02413 and No.~UMO-2023/49/N/ST3/03032.
\end{acknowledgments}

\section*{Author Contributions}
M.G., K.S. and M.K. planned and executed the study. Contributions are as follows: M.G.: conceptualization, methodology, software, formal analysis, investigation, data curation, writing - original draft, visualization, funding acquisition. K.S.: methodology, validation, formal analysis, investigation, writing - review \& editing. M.K.: conceptualization, validation, resources, writing - review \& editing, supervision, project administration, funding acquisition.

\section*{Competing Interests}
The authors declare no competing interests.

\section*{Additional Information}
Supplementary Information is available for this paper at \textbf{TBA}.

\end{document}